\documentclass[final]{vgtc}                          % final (conference style)

\graphicspath{{figures/}{pictures/}{images/}{./}} % where to search for the images

\usepackage{times}                     % we use Times as the main font
\usepackage{tabu}                      % only used for the table example
\usepackage{booktabs}                  % only used for the table example
\usepackage{lipsum}                    % used to generate placeholder text
\usepackage{mwe}                       % used to generate placeholder figures
\usepackage{mathptmx}                  % use matching math font
\usepackage{amsmath}
\onlineid{1263}
\vgtccategory{Poster}
\vgtcinsertpkg

\title{ClipGS-VR: Immersive and Interactive Cinematic Visualization of
Volumetric Medical Data in Mobile Virtual Reality}

\author{Yuqi Tong, Ruiyang Li\thanks{Corresponding author}, Chengkun Li, Qixuan Liu, Shi Qiu, Pheng-Ann Heng\\ %
        \scriptsize Department of Computer Science and Engineering and Institute of Medical Intelligence and XR, The Chinese University of Hong Kong\\
        \scriptsize \{yqtong, liry, ckli, qxliu, shiqiu,  pheng\}@cse.cuhk.edu.hk}

\abstract{
  High-fidelity cinematic medical visualization on mobile virtual reality (VR) remains challenging. Although ClipGS enables cross-sectional exploration via 3D Gaussian Splatting, it lacks arbitrary-angle slicing on consumer-grade VR headsets. To achieve real-time interactive performance, we introduce ClipGS-VR and restructure ClipGS's neural inference into a consolidated dataset, integrating high-fidelity layers from multiple pre-computed slicing states into a unified rendering structure. Our framework further supports arbitrary-angle slicing via gradient-based opacity modulation for smooth, visually coherent rendering. Evaluations confirm our approach maintains visual fidelity comparable to offline results while offering superior usability and interaction efficiency.
} % end of abstract


\begin{document}


\firstsection{Introduction}

\maketitle

%% \section{Introduction} %for journal use above \firstsection{..} instead
Medical visualization enables clinicians to gain an intuitive understanding of complex spatial relationships, transforming raw volumetric data into actionable diagnostic insights. Recent advances in 3D Gaussian Splatting (3DGS) \cite{kerbl20233d} have made real-time rendering of volumetric medical data feasible; for example, Kleinbeck \emph{et al.} utilized multi-layer Gaussian training to display distinct anatomical structures \cite{kleinbeck2025multi}. However, existing methods often represent anatomy as hollow surfaces or restrict users to pre-defined layers, and thus fail to support flexible medical data exploration. 

To address this limitation, ClipGS \cite{li2025clipgs} introduces a learnable truncation strategy that explicitly models the solid nature of tissue and enables cross-sectional slicing, rendering both internal textures and external surfaces and thereby achieving high-quality volumetric reality by leveraging the latest neural rendering technique. However, ClipGS allows only uniaxial cutting-plane transformations, with no support for arbitrary slicing. In addition, its substantial computational overhead (exceeding 6GB of VRAM) typically limits deployment to high-end workstations, restricting real-world deployment on consumer-grade immersive devices.

In this work, we present a novel mobile framework that adapts ClipGS for immersive applications and extends its capabilities to support arbitrary-angle slicing. Our framework introduces a generalized arbitrary slicing mechanism, synthesizing coherent views from any orientation via gradient-based opacity modulation. By consolidating the highest-fidelity layers from discrete precomputed states into a unified, lightweight structure, we avoid runtime neural inference and preserve the cinematic fidelity of ClipGS within the strict performance and memory constraints of mobile VR, thereby enabling interactive, high-resolution anatomical slicing.

\section{Method}

\subsection{Data Preprocessing and Compression}

To adapt the ClipGS framework for realtime rendering on standalone mobile VR, we develop a data preprocessing pipeline that eliminates the need for runtime neural network inference. The original ClipGS architecture relies on the Learnable Truncation (LT) scheme and the Adaptive Adjustment Model (AAM) to dynamically manipulate Gaussian visibility and deformation during slicing \cite{li2025clipgs}. As real-time execution of these modules on mobile VR platforms is infeasible, we replace continuous inference with a discrete pre-computation strategy. We uniformly sample 200 slicing plane states across the volumetric extent of each case in the ClipGS dataset and calculate the optimized outputs of both LT and AAM for each state. The resulting neural module outputs are then baked directly into the Gaussian attributes, converting the implicit neural processing into explicit 3D Gaussian representations. 

\begin{figure}[tb]
 \centering % avoid the use of \begin{center}...\end{center} and use \centering instead (more compact)
 \includegraphics[width=\columnwidth]{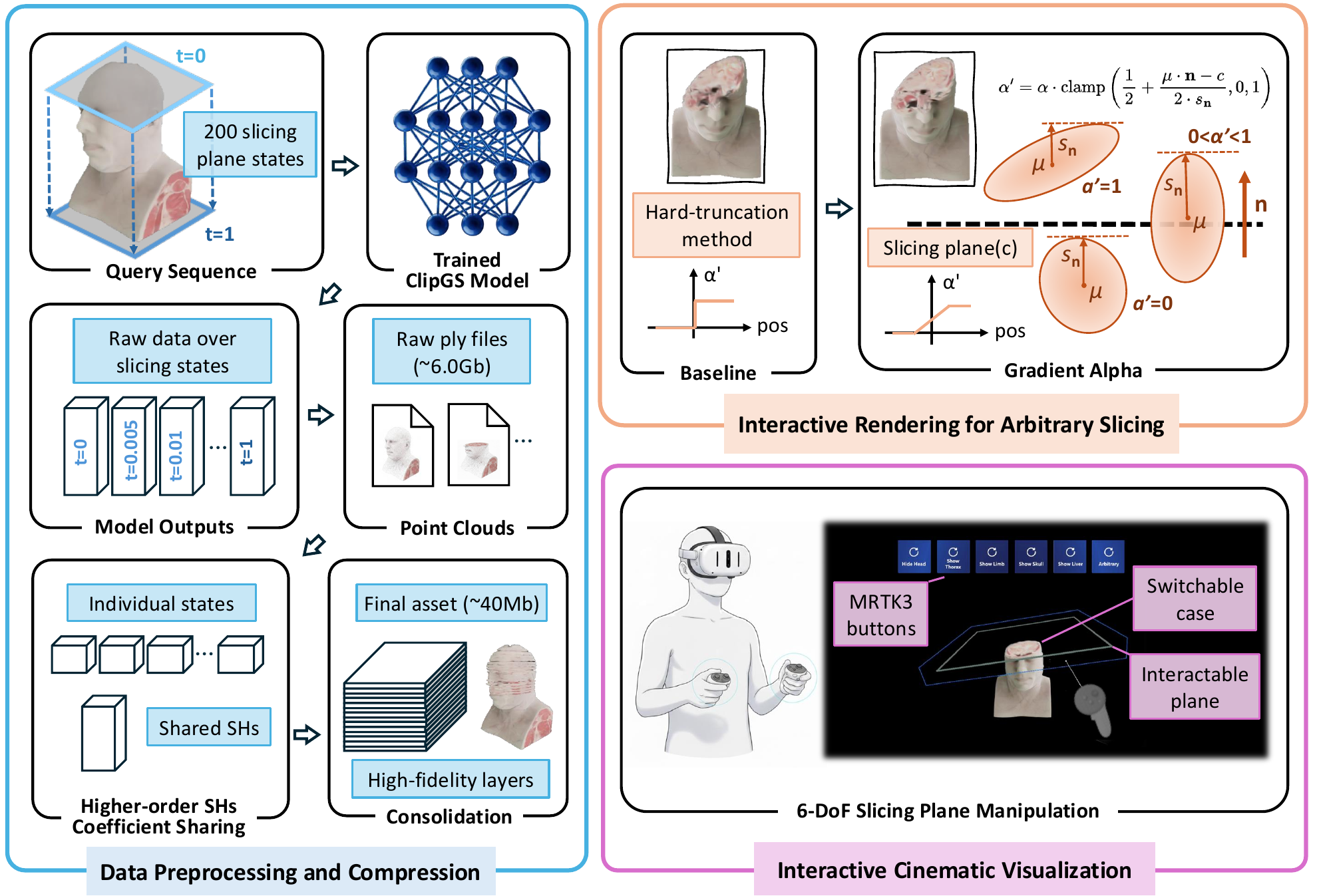}
 \caption{The overall workflow of the proposed method ClipGS-VR.}
 \label{fig:overview}
\end{figure}
% The overall workflow of the proposed method ClipGS-VR, including data preprocessing and compression, and interactive rendering for arbitrary slicing. The system also achieves interactive cinematic visualization for volumetric medical data in mobile VR.

Following the generation of these discrete states, we utilize the Unity Gaussian Splatting framework\footnote{\href{https://github.com/aras-p/UnityGaussianSplatting}{https://github.com/aras-p/UnityGaussianSplatting}.} to serialize the point cloud data of 3D Gaussians into a compact binary asset format. We incorporate two critical compression strategies to minimize memory overhead. First, we find that the ClipGS network preserves spherical harmonics coefficients of Gaussian points beyond the zeroth order. Hence, we decouple the higher-order coefficients from the individual states and store a single global reference array of spherical harmonics coefficients that is shared across all 200 states. Second, we reduce spatial redundancy in sequential slicing, where geometric differences between adjacent states are confined to the immediate vicinity of the cutting plane. Instead of duplicating the entire volume of each state, we retain only the distinct primitives at the slicing plane, which are the highest-fidelity ones at the horizontal layer. We then consolidate all point clouds into a unified structure, assigning each slicing state to a specific rendering layer to facilitate efficient runtime transitions. Our approach reduces each sequence of raw point cloud PLY files ($\sim$6.0 GB per case) to a final asset averaging only $\sim$40 MB per case.

\subsection{Interactive Rendering for Arbitrary Slicing}

To overcome the limitation of uniaxial cutting-plane transformation in the original ClipGS framework, we extend the rendering pipeline to support arbitrary slicing orientations for interactive exploration. A naive baseline implementation provides a fundamental cutout mechanism, which relies on a binary visibility test that renders primitives solely based on whether their centroids reside on the positive side of the slicing plane. However, this hard truncation ignores the volumetric extent of the splats, inevitably resulting in jagged aliasing artifacts and visual discontinuities along the cutting boundary. To resolve these artifacts and achieve a high-fidelity rendering, we introduce a gradient-based opacity modulation. Specifically, for any rendering primitive centered at $\mathbf{\mu}$, we adjust its original opacity $\alpha$ by calculating a transparency coefficient $\sigma$ related to the ratio of the primitive's signed distance to the cutting plane $c$ and its projected radius $s_{\mathbf{n}}$ along the plane's normal vector $\mathbf{n}$. If $\sigma$ is within (0,1), it is used to modify the final opacity $\alpha'$ as: 
$\alpha' = \alpha \cdot \sigma \text{, where } \sigma = \text{clamp}\left( 1/2 + (\mathbf{\mu} \cdot \mathbf{n} - c)/(2 \cdot s_{\mathbf{n}}), 0, 1 \right)$.
% \begin{equation}
%     \alpha' = \alpha \cdot \sigma \text{, where } \sigma = \text{clamp}\left( \frac{1}{2} + \frac{\mathbf{\mu} \cdot \mathbf{n} - c}{2 \cdot s_{\mathbf{n}}}, 0, 1 \right)
% \end{equation}
% \begin{equation}
%     \alpha' = \alpha \cdot \text{clamp}\left( \frac{1}{2} + \frac{\mathbf{\mu} \cdot \mathbf{n} - c}{2 \cdot s_{\mathbf{n}}}, 0, 1 \right)
% \end{equation}
This modulation ensures that Gaussians intersecting the slicing plane fade out smoothly according to their geometric penetration depth, while maintaining high rendering speed.

\subsection{Implementation} 

We implement the interaction logic using the Mixed Reality Toolkit 3 (MRTK3)\footnote{\href{https://github.com/MixedRealityToolkit/MixedRealityToolkit-Unity}{https://github.com/MixedRealityToolkit/MixedRealityToolkit-Unity}.}. The core mechanism is a 6-degrees-of-freedom (6-DoF) slicing plane that users manipulate directly via VR controllers. The transform of this virtual plane is streamed in real time to the rendering shader, enabling precise, arbitrary slicing of the anatomy. In addition, we provide a menu interface with buttons for users to rapidly switch between clinical cases. All experiments are conducted on a standalone Meta Quest 3\footnote{\href{https://www.meta.com/quest/quest-3/}{https://www.meta.com/quest/quest-3/}.} headset.

\section{Evaluation}

% We evaluated the proposed framework from two complementary perspectives. First, we assessed the visual fidelity of our mobile implementation against the baseline hard-truncation method. Second, we conducted a user study employing the System Usability Scale (SUS) and interaction efficiency to compare the original uniaxial slicing versus our proposed arbitrary slicing mechanism.

\subsection{Visual Comparison} 

\begin{figure}[t]
 \centering % avoid the use of \begin{center}...\end{center} and use \centering instead (more compact)
 \includegraphics[width=\columnwidth]{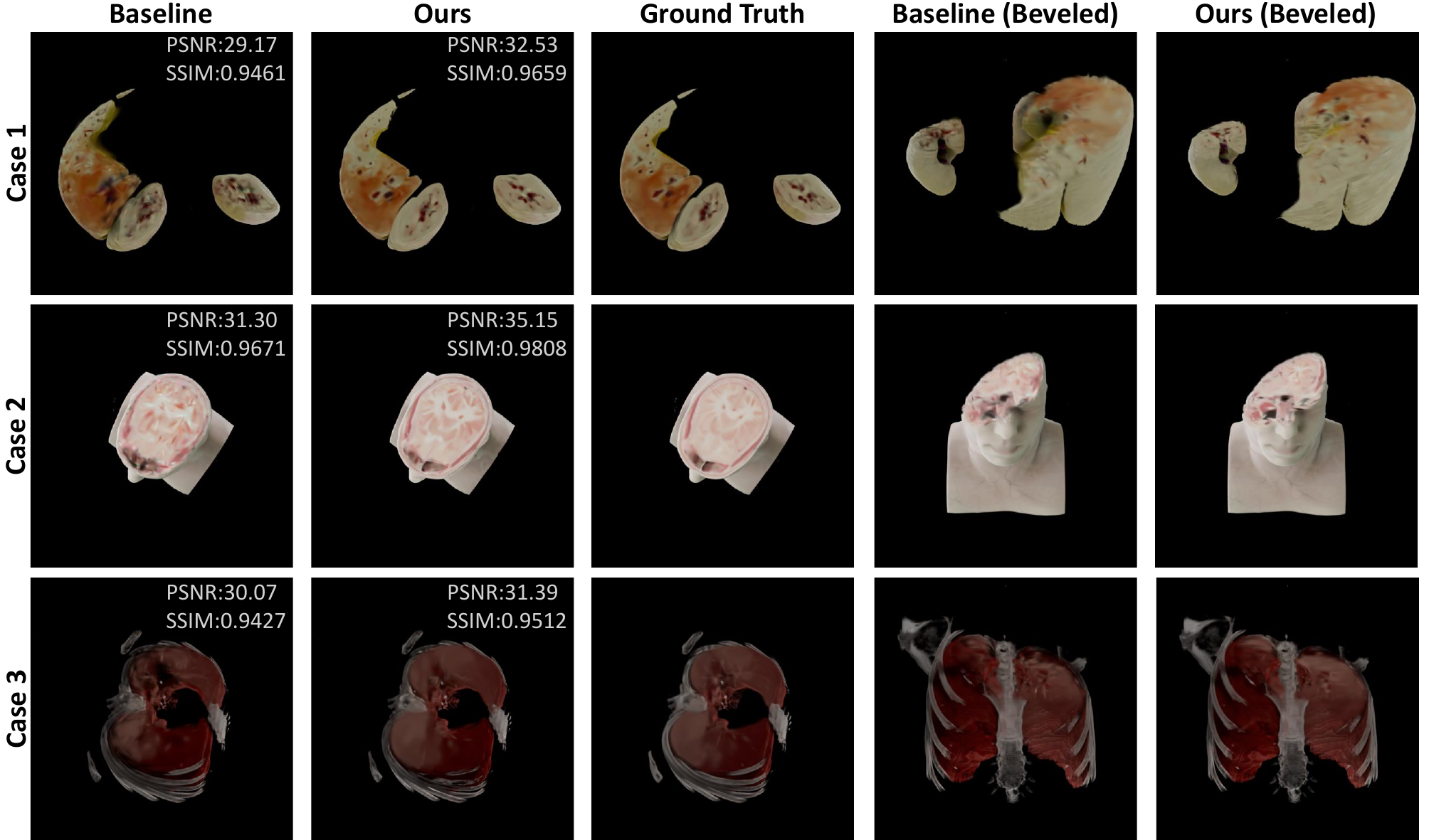}
 \caption{Visual comparison under uniaxial slicing mode (columns 1-3) and beveled slicing mode (columns 4-5).}
 \label{fig:comp}
\end{figure}

We compared our rendering output against the baseline hard-truncation method with the ground truth generated by the original ClipGS method. For quantitative analysis, we reported the metrics of Peak Signal-to-Noise Ratio (PSNR) and Structural Similarity Index (SSIM) between the rendered image and the ground truth under uniaxial slicing mode. As shown in the cases 1-3 in Fig.~\ref{fig:comp}, our method achieves an average PSNR of $33.40$ dB and SSIM of $0.9698$ versus the baseline method with PSNR of $30.55$ dB and SSIM of $0.9542$, indicating our better rendering quality. For arbitrary slicing settings without ground-truth references, we provide a qualitative comparison in columns 4-5 in Fig.~\ref{fig:comp}. When the slicing plane intersects the volume at beveled angles, our method produces smooth, anti-aliased cutting surfaces, effectively mitigating the jagged artifacts observed in the baseline implementation.

\subsection{User Study}

We conducted a user study with 10 participants tasked with inspecting anatomical structures under two interaction modes: a fixed uniaxial slicing, and our proposed method of arbitrary slicing. 
Results on System Usability Scale (SUS) scores and interaction efficiency with a 5-point Likert scale are shown in Tab.~\ref{tab:sus_comparison}. A paired t-test revealed a significantly higher SUS of our arbitrary slicing approach than that of the uniaxial slicing ($t(9)=-2.988,p=0.015$). We analyzed the non-normality scores on interaction efficiency with a Wilcoxon signed-rank test and found a significant difference between interaction methods ($W=0.00,Z=-2.739,p=0.006$). Moreover, participants reported that the uniaxial slicing mode often requires uncomfortable physical postures to view specific angles, whereas our arbitrary slicing method allows natural alignment with internal structures. These findings indicate that full 6-DoF interaction is critical for reducing physical fatigue and enhancing spatial understanding in immersive medical visualization.

% Visual comparison under different slicing cases and angles. Columns 1–3 show the PSNR/SSIM scores of our arbitrary slicing method compared with the baseline method under uniaxial slicing mode. Columns 4–5 illustrate beveled slicing effects for our method and the baseline.

\begin{table}[t]
  \caption{Comparison on SUS and interaction efficiency scores (Mean $\pm$ SD) for uniaxial slicing and arbitrary slicing.}
  \label{tab:sus_comparison}
  \scriptsize%
  \centering%
  % 使用 to \linewidth 确保在单栏内部左右顶格
  \begin{tabu} to \linewidth {
    X[2.5,l]  % Method Name
    X[2,c]  % SUS Score
    X[2.5,c]  % Satisfaction
    }
  \toprule
    Interaction Mode & SUS Score & Interaction Efficiency \\
  \midrule
    Uniaxial Slicing & $71.80 \pm 14.41$ & $3.40 \pm 0.70$ \\
    Arbitrary Slicing & $88.20 \pm 9.35$ & $4.70 \pm 0.48$ \\
  \bottomrule
  \end{tabu}%
\end{table}

\section{Conclusion}

In this work, we present a novel mobile framework adapting ClipGS for immersive medical visualization with interactive arbitrary slicing on standalone VR headsets. By consolidating the highest-fidelity layers from discrete pre-computed states into a unified rendering structure, our system achieves real-time performance without runtime neural inference. We further extend the framework to support arbitrary slicing orientations using gradient-based opacity modulation. Comprehensive evaluations show that our approach maintains high visual fidelity while improving system usability and interaction efficiency. Future work will focus on optimizing the precomputation pipeline and exploring real-time deformation to support dynamic medical simulation scenarios.

\acknowledgments{This work was supported by the Research Grants Council of the Hong Kong Special Administrative Region, China (Project No.: T45-401/22-N); and
in part by The Chinese University of Hong Kong (Project No.: 4055212).}

\bibliographystyle{abbrv-doi}

\bibliography{ref}
\end{document}